\documentclass[aps,twocolumn,amsmath,amssymb,floatfix,superscriptaddress]{revtex4}

\usepackage{hyperref}
\usepackage{amsmath}
\usepackage[dvips]{graphicx}
\usepackage{epsfig}
\usepackage{amssymb}
\usepackage{amsfonts}
\usepackage{tabularx}
\usepackage{graphicx}
\usepackage{color}
\usepackage{natbib}

\begin{document}

\title{Slow thermo-optomechanical pulsations in suspended 1D photonic crystal nanocavities}

\author{Piergiacomo Z.G. Fonseca}\email{piergiacomo.fonseca@icfo.eu}
\affiliation{Institut de Ciencies Fotoniques, Mediterranean Technology Park, 08860 Castelldefels, Barcelona, Spain}

\author{Irene Alda}
\affiliation{Institut de Ciencies Fotoniques, Mediterranean Technology Park, 08860 Castelldefels, Barcelona, Spain}

\author{Francesco Marino}
\affiliation{Istituto Nazionale di Ottica, Via Sansone 1, I-50019 Sesto Fiorentino, Florence, Italy}

\author{Alexander Cuadrado}
\affiliation{Escuela de Ciencias Experimentales y Tecnología, University Rey Juan Carlos, Móstoles, 28933, Madrid, Spain}

\author{Vincenzo d'Ambrosio}
\affiliation{Institut de Ciencies Fotoniques, Mediterranean Technology Park, 08860 Castelldefels, Barcelona, Spain}
\affiliation{Dipartimento di Fisica, Università di Napoli Federico II, Complesso Universitario di Monte S. Angelo, Via Cintia, 80126 Napoli, Italy}

\author{Jan Gieseler}
\affiliation{Institut de Ciencies Fotoniques, Mediterranean Technology Park, 08860 Castelldefels, Barcelona, Spain}

\author{Romain Quidant}\email{rquidant@ethz.ch}
\affiliation{Institut de Ciencies Fotoniques, Mediterranean Technology Park, 08860 Castelldefels, Barcelona, Spain}
\affiliation{Institució Catalana de Recerca i Estudis Avançats, 08010 Barcelona, Spain}
\affiliation{Nanophotonic Systems Laboratory, Department of Mechanical and Process Engineering, ETH Zurich, 8092 Zurich, Switzerland}

\date{\today}
\begin{abstract}
We investigate the nonlinear optical response of suspended 1D photonic crystal nanocavities fabricated on a silicon nitride chip. Strong thermo-optical nonlinearities are demonstrated for input powers as low as $2\,\mu\text{W}$ and a self-sustained pulsing regime is shown to emerge with periodicity of several seconds. As the input power and laser wavelength are varied the temporal patterns change in period, duty cycle and shape. This dynamics is attributed to the multiple timescale competition between thermo-optical and thermo-optomechanical effects and closely resembles the relaxation oscillations states found in mathematical models of neuronal activity. We introduce a simplified model that reproduces all the experimental observations and allows us to explain them in terms of the properties of a 1D critical manifold which governs the slow evolution of the system.
\end{abstract}

\maketitle

\section{Introduction} 

Advances in the field of integrated optics rely on the development of compact optical components where light can be tightly confined and processed within a chip-scale photonic structure \cite{Soref2006,Rahim2017,Atabaki2018}. Adversely, one of the results of successfully localizing light at the sub-micron scale is the strong field enhancement within the photonic component, which may lead to nonlinear optical effects even at moderate input powers. While these effects can lead to deleterious device instabilities, they also offer new opportunities for all-optical sensing and low-power signal processing applications. \cite{Wang2012}. For this reason, optical nonlinear phenomena in micro- and nano-photonic devices have been the object of extensive investigations for many years \cite{Soref1987,Barclay2005,Priem2005}.

Due to its potential applications in all-optical memories, switching and logic gates, optical bistability \cite{Gibbs1985} received considerable attention. Bistable behavior coming from different nonlinear mechanisms has been reported for a vast number of micro- and nano-optical resonators \cite{Almeida2004,Xu2006,Johnson2006,Zhang2014}, whose combination of high quality factors $Q$ and small mode volumes $V_\text{m}$ allows to achieve high intra-cavity optical power densities even at very low input powers. Photonics crystal (PhC) nanocavities, either in one (1D) or two dimensions (2D), have demonstrated incresingly low bistability thresholds \cite{Notomi2005,Tanabe2005,Weidner2007,Jiang2019}, until achieving values as low as $\sim2\,\mu\text{W}$ \cite{Weidner2007,Haret2009}.

In the presence of competing nonlinearities operating at very different timescales, the bistability often breaks down and the system enters a regime of self-sustained pulsations (SSPs), also known in dynamical systems theory as \emph{relaxation oscillations} \cite{ROreview}. The evolution is characterized by periods of slow motion separated by faster relaxation jumps between them, which result in a sequence of square-wave-like pulses. This characteristic pattern is generally determined by the multiple timescale competition between two effects, with the slower driving the system across the hysteresis cycle induced by the faster. 

\begin{figure*}
\begin{center}
\includegraphics*[width=1.8\columnwidth]{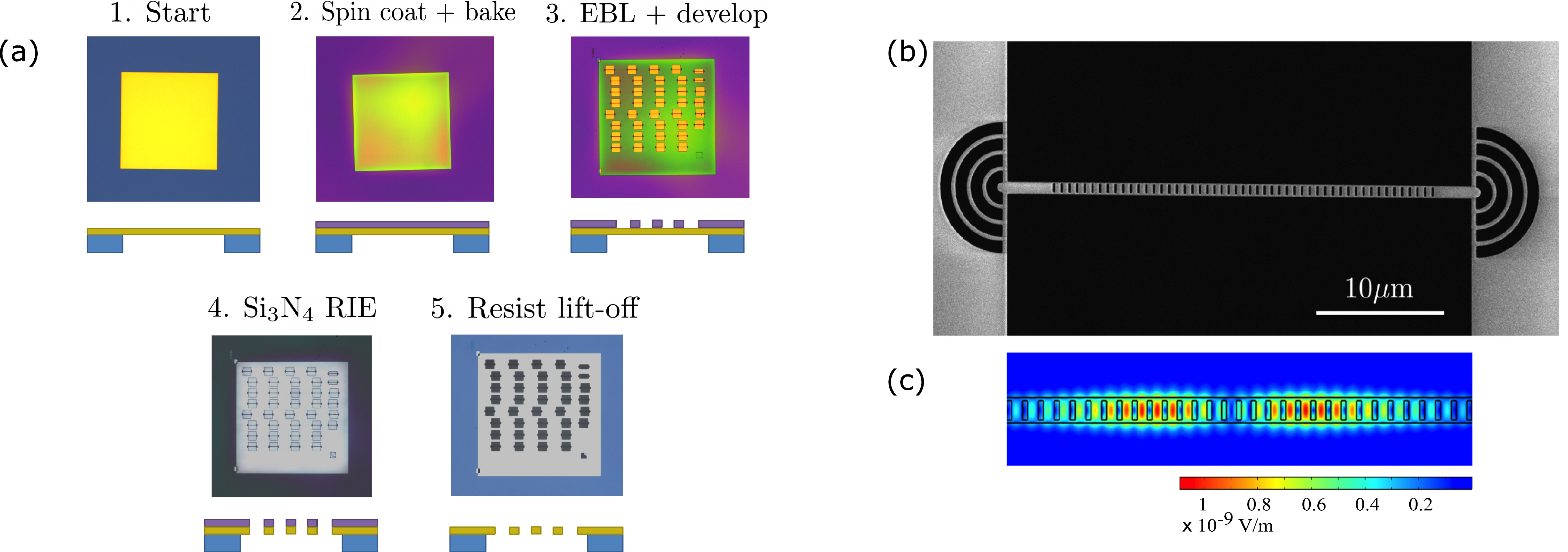}
\caption{Schematic processing steps and corresponding optical microscope images of the sample. 1) The Si$_3$N$_4$ membrane before any fabrication step. 2) The sample after spin coating of a $\sim300\,\text{nm}$-thick film of resist (CSAR 62) and 1 minute baking to stabilize the resist. 3) After exposure to an electron beam at $30\,\text{kV}$ and $166\,\mu\text{A}$ of emission current for 1 h. The sample is then introduced into a developer (AR600-546) and a stopper (IPA), to remove the exposed regions and provide a mask for the following etching procedure. 4) The sample after 10 minutes RIE of the Si$_3$N$_4$ performed with O$_2$ and CHF$_3$ gases. 5) After 1 minute O$_2$ cleaning process for resist lift-off. (b) SEM image of the PhC nanocavity with the Bragg grating couplers. (c) COMSOL simulation of the electric field profile at resonance.}
\label{figure1}
\end{center}
\end{figure*}

In the case of micro- and nano-cavities, the above dynamical mechanism manifests itself in the alternating shift of the cavity resonance to opposite directions, with characteristic frequencies that depend on the physical processes involved. SSPs ranging between the kHz- to the MHz-scale have been observed, e.g. in semiconductor micro-cavities \cite{barland2003,marino2004,marino2005} and photonic crystals \cite{yacomotti2006} due to the interplay between carrier-induced and thermo-optic (TO) nonlinearities, and in silica \cite{Ilchenko1992,Fomin2005,Park2007} and polymer \cite{Luo2014} micro-resonators, due to competing Kerr and TO effects. Faster oscillations from a few MHz up to the GHz-scale have been reported in silicon micro- and nano-cavities, where they originate from the nonlinear coupling between optical, free-carrier and thermal variables \cite{Johnson2006,Zhang2014,Yang2014,Navarro2015}, and in silica toroidal microcavities induced by the interplay between radiation-pressure and the intra-cavity field \cite{Carmon2005}. On the other hand, when thermal expansion processes come into play, SSPs can become extremely slow with periods of a few seconds, as observed e.g. in silicon nitride microdisks \cite{Baker2012} and calcium fluoride WGM resonators \cite{Weng2015}. 

Different applications for systems displaying SSPs have been proposed, for instance in continuous pulse laser generation \cite{He2008} and sensing \cite{Deng2013}. More recently, a new research area, known as neuromorphic photonics, has emerged aiming at establishing a bridge between photonic devices and neural networks \cite{NPreview}. In this context, photonic systems displaying SSPs are the key-elements in view of realizing "brain-inspired" computing platforms and/or simulating complex neuron dynamics.

In this work, we introduce a novel 1D suspended PhC nanocavity device, fabricated on a free-standing Si$_3$N$_4$ thin membrane. Its simple and compact on-chip geometry ensures full scalability, while free-space optical access provides an easy opportunity for parallel signal processing. As in similar designs  \cite{Haret2009,Navarro2015}, the suspended configuration limits heat dissipation and favors nonlinear effects due to strong thermo-optic confinement. As a consequence, the nanocavity exhibits strong TO effects at injected powers as low as $\sim2\,\mu\text{W}$. When the laser is detuned to the red-side of the cavity resonance, periodic SSPs are observed with sub-Hz characteristic frequencies. Such a slow periodicity allows us to exclude a number of nonlinear effects such as Kerr, carrier-induced and radiation-pressure, and to attribute SSPs to the interaction between a faster TO effect, and a slow thermo-optomechanical (TM) mechanism. On this basis we construct a simplified model that reproduces all the observed phenomenology and fits well the experimental data. We finally show that the phase-space structure of the system is equivalent to that of the Van der Pol–-FitzHugh-Nagumo (VdPFN) neuron model \cite{fitgh1961impulses,nagumo1962active,fhnbook}, where SSPs results from the existence of a 1D critical manifold which organizes the dynamics on a slow timescale.

The paper is organized as follows. In Sec. II we describe the PhC nanocavity fabrication process and the experimental apparatus. In Sec. III we show the cavity transmission spectra obtained when the laser frequency is scanned downwards across a single mode of the resonator. At higher input powers the spectra become highly nonlinear providing a clear evidence of TO effect. In Sec. IV we theorize about the TM mechanism at the basis of our observations and introduce a simplified physical model. In Sec. V we present the experimental results on the SSP dynamics and quantitatively compare them with the numerical predictions. In Sec. VI we analyze the bifurcations of the model and explain the emergence of SSPs in our system by means of geometric singular perturbation theory. Conclusions and future perspectives are reported in Sec. VI.

\section{PHC Nanocavity and Experimental Setup}

The 1D Si$_3$N$_4$ PhC nanocavities are fabricated starting from an amorphous $0.5\times0.5 \,\rm{mm^2}$ Si$_3$N$_4$ membrane (Norcada), with thickness $h = 200$ nm. The steps of the fabrication process are summarized in Fig. \ref{figure1}(a) and include an electron-beam lithography procedure to pattern the desired structures on the membrane, followed by a sequence of reactive ion etching (RIE) and O$_2$ plasma lift-off, to etch the Si$_3$N$_4$ and remove any left-over resist. Fabrication is highly scalable, as we can fit together up to 40 nanocavities within a single membrane. 

The optical resonator is formed by a suspended nanobeam of length $l=38\,\mu\text{m}$ and width $w=1\,\mu\text{m}$, containing a periodic array of rectangular air-holes that create the photonic bandgap (see Fig \ref{figure1}(b)). The width of the two bridges above and below the air-holes is $e\simeq130\,\text{nm}$. 
To confine light and obtain a cavity mode, a defect is tailored where the periodicity is quadratically reduced from the sides to the center of the structure. The complete design includes 20 unit-cells of constant periodicity on each side (mirror cells) and 19 defect cells at the center of the nanobeam. Light is coupled in and out of the nanocavity through two curved Bragg-like grating couplers \cite{Barth2015,Zhu2017} (see figure \ref{figure1}(b)). A COMSOL simulation of such a design reveals that the 2nd order cavity mode is the dominant optical mode to be confined within the bandgap, as shown in Fig. \ref{figure1}(c). 

\begin{figure}
\begin{center}
\includegraphics*[width=0.45\textwidth]{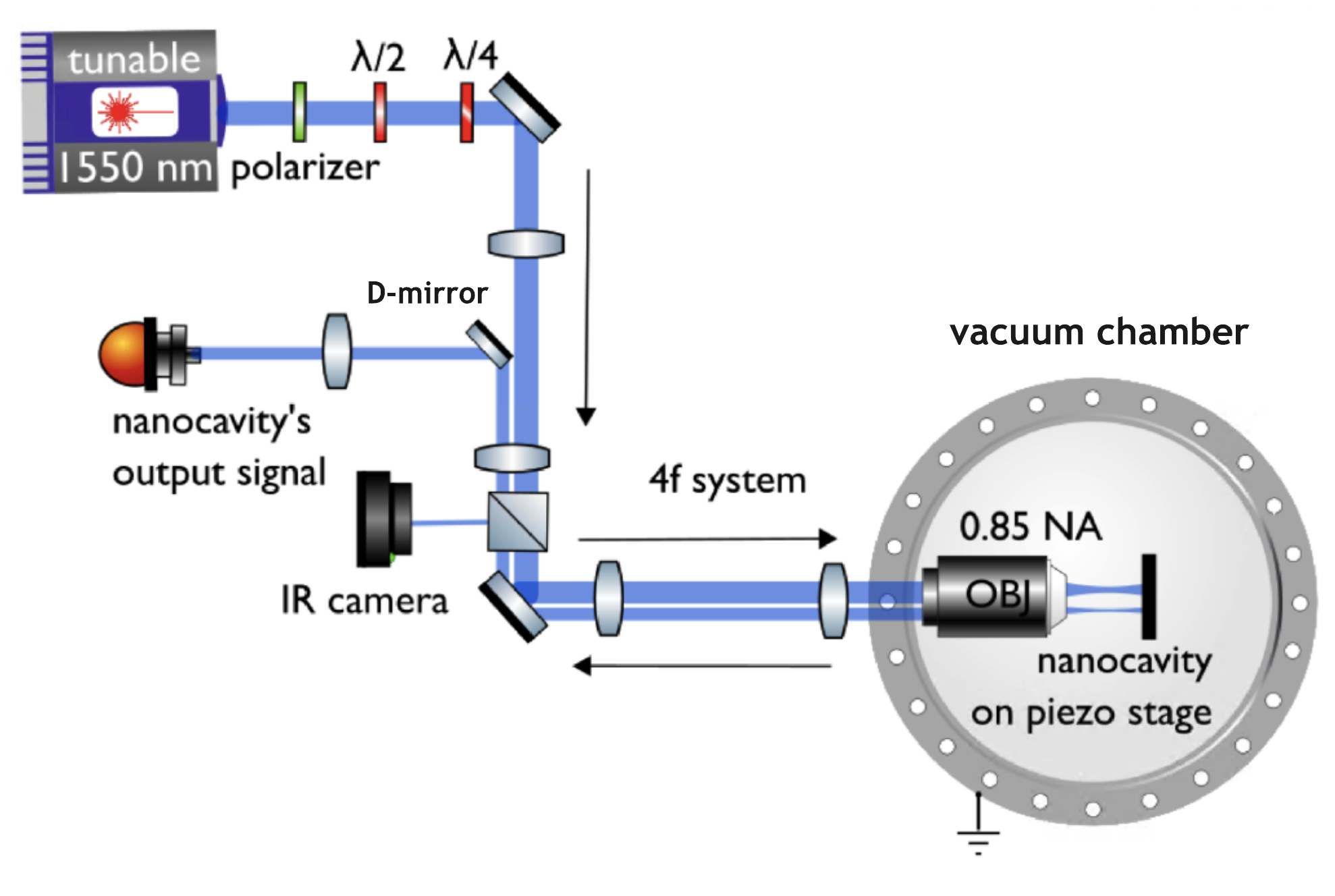}
\setlength{\belowcaptionskip}{-29.8pt}
\caption{Schematic of the experimental setup. Light is coupled in and out of the nanocavity via a high numerical aperture microscope objective. The input light is linearly polarized to minimize waveguide propagation losses. A $4f$-system is used to image the nanocavity onto an IR camera, and a D-shaped mirror allows to only detect the output light from the cavity.}.
\label{figure2}
\end{center}
\end{figure}

A scheme of the optical setup is reported in Fig. \ref{figure2}. Light from a tunable external-cavity laser (Tunics Plus) at $1550\,\text{nm}$ is injected into the nanocavity via a high-Numerical-Aperture (NA=0.85) microscope objective. Both the nanocavity and the objective are mounted inside a vacuum chamber, evacuated at $p\simeq0.5\,\text{mbar}$, and the sample sits on a 3D piezo translational stage to optimize optical alignment. The light transmitted by the cavity and scattered by the second coupler is collected by the same objective and sent to detection. We use a $4f$-system of lenses to separate the cavity transmitted signal from the input laser light before the photoreceiver, and to image the nanocavity onto an IR camera. To estimate the coupling efficiency, we use non-structured waveguide nanobeams fabricated on the same membrane, and compare the amount of power going in from the objective, with the final power detected by the photodiode. Taking into account the different losses channels involved in the system, we estimate that $\simeq10\%$ of the light incident on the input coupler is injected into the nanobeam. 

\section{Competing thermal nonlinearities}

We first characterize the system response, as both the laser frequency and injected power are varied. Since the laser is not frequency-locked, drifts of the cavity resonance and/or of the laser wavelenght and power, prevent the construction of the spectrum by adiabatically tuning the laser frequency. On the other hand, the characteristic nonlinear features of trasmission spectra could be smoothened at high scanning rates (of the order of frequency cut-off of the nonlinear effect, or faster). 
In our measurements we use a scanning rate of 5 pm/s, sufficiently fast to avoid the effect of possible long-term changes of the intra-cavity field and keep the same experimental conditions, but slower than the expected TO response in our system.

\begin{figure*}
\begin{center}
\includegraphics[width=0.99\textwidth]{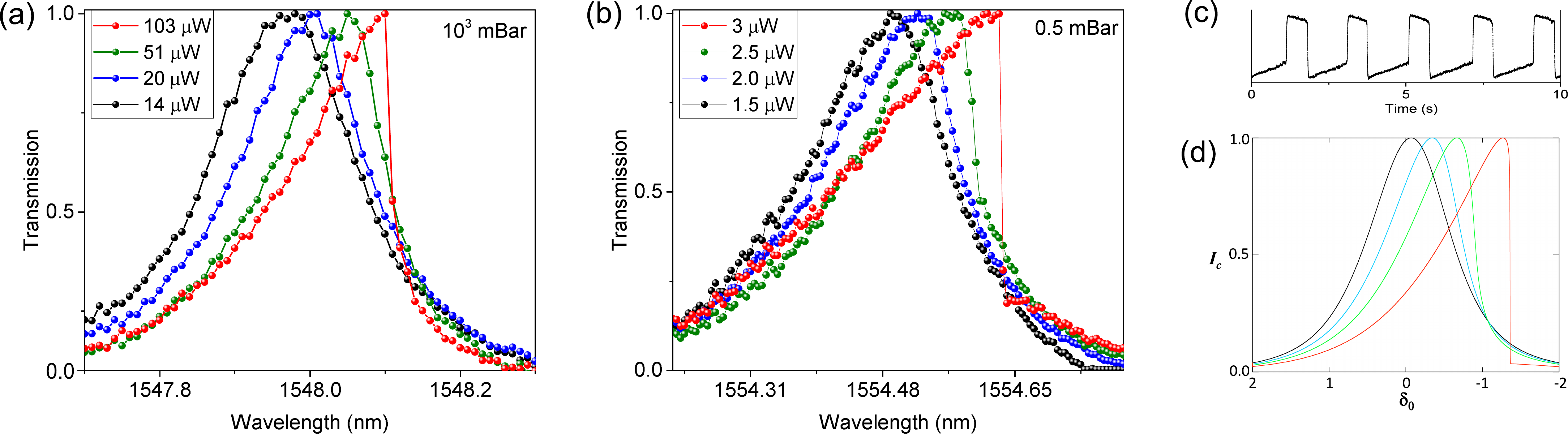}
\caption{Cavity transmission spectra at different input powers at (a) atmospheric pressure and (b) $p\simeq0.5\,\text{mbar}$, taken by scanning the laser wavelength across resonance. (c) Time trace of the transmission intensity for input power $P_\text{in}=2.25\,\mu\text{W}$ and detuning $\delta=\lambda_\text{res}-\lambda_\text{L}=-0.13\,\text{nm}$. (d) Transmission spectra as obtained by numerical integration of Eqs. (\ref{eq1}-\ref{eq3}) by tuning the detuning parameter $\delta_0$ in the interval [$2$,$-2$] (from shorter to longer wavelenghts) with a scan rate $\varepsilon=10^{-3}$.}
\label{figure3}
\end{center}
\end{figure*}

In Fig. \ref{figure3} we show the transmitted intensity as the laser is swept from shorter to longer wavelengths across the cavity resonance at (a) atmospheric pressure and (b) at $p \sim 0.5\,\text{mbar}$. For clarity, the intensity signals are normalized to be zero away from resonance and equal to 1 at resonance. For the lowest power used, the system still operates in the linear regime showing a typical Lorentzian shape. From this spectrum we estimate a cavity half-linewidth $\gamma\simeq0.12\,\text{nm}$, corresponding to a cavity quality factor $Q=\lambda_\text{res}/2\gamma\simeq6500$, where $\lambda_\text{res} \sim 1.55\,\mu\text{m}$ is the cavity resonant wavelength. When the input power is increased, $\lambda_\text{res}$ is red-shifted, as expected in Si$_3$N$_4$ owing to a positive TO coefficient. The thermal origin of the nonlinearity is further supported by the observation that the threshold power to enter the nonlinear regime decreases as $p$ is lowered from atmospheric pressure, due to the reduced contribution of convective heat dissipation (see spectra in Fig. \ref{figure3}(a,b)). At higher intensities and low-pressure, the spectral line takes an asymmetric saw-tooth profile, with a sharp drop on the red-side of the resonance. While similar spectral shapes are often interpreted as the signature of optical bistability \cite{Notomi2005,Haret2009}, they can also arise in the presence of dynamical instabilities if the scanning rate of the laser wavelength is of the order of (or faster than) the instability growth rate. 
When the input power $P_\text{in}$ is kept constant and the laser wavelength $\lambda_\text{L}$ is fixed and red-detuned with respect to the cavity resonance, the system enters a self-sustained oscillatory regime (see Fig. \ref{figure3}(c)). The time-series, consisting of a periodic sequence of fast switchings between slowly evolving high- and low-transmission states, display the characteristic pattern of relaxation oscillations \cite{ROreview}. These dynamics imply the existence of a second nonlinear effect, evolving on a slower timescale and providing an opposite shift to the cavity resonance with respect to the TO.

The slow timescale of the process (a few seconds) allows us to rule out most of the typical nonlinear effects in PhC suspended nanocavities such as Kerr, carrier-induced or radiation-pressure and suggests a thermo-optomechanical mechanism \cite{Weng2015,Baker2012}. The thermal expansion coefficient of Si$_3$N$_4$ is positive: in this case, the simple expansion of the material and deformation of the structure would induce a red-shift of the resonance. On the other hand, local heating in PhC suspended nanocavities gives rise also to complex buckling of the nanobeam which may result in a slow blue-shifting optical nonlinearity \cite{Baker2012}. For instance, we checked that a simple inward bending at the center of the structure induces a blue-shift of the cavity resonance. 

FEM simulation are required for a quantitative assesment of both the thermo-optomechanical and the TO effect, to account for thermally-induced stress in the SiN nanonbeam and to model the spatial extent of the absorbing region, which strongly depends upon the geometry of the device.
Here instead, we are interested in deriving a simple ordinary differential equation (ODE) model, that is able to reproduce the observations and to identify the undelying mechanisms at the basis of SSPs in our system. This is what we discuss in the next section.

\section{Physical model}

\begin{figure*}
\begin{center}
\includegraphics[width=1.8\columnwidth]{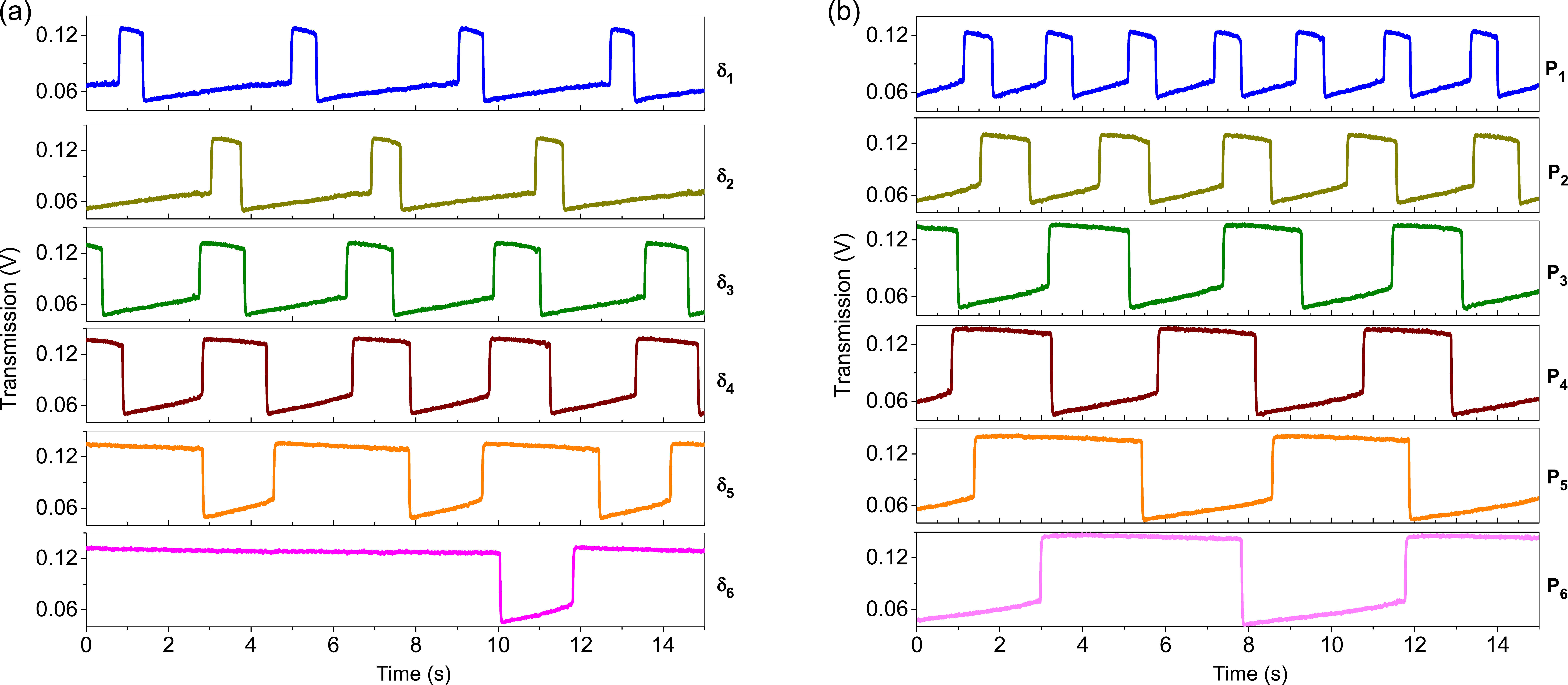}
\caption{Time-traces of cavity transmission signal in the SSP regime: (a) fixed input power $P_\text{in}=2.5\,\mu\text{W}$ and detunings $\delta_1=-0.14\,\text{nm}$, $\delta_2=-0.12\,\text{nm}$, $\delta_3=-0.11\,\text{nm}$, $\delta_4=-0.10\,\text{nm}$, $\delta_5=-0.09\,\text{nm}$, $\delta_6=-0.08\,\text{nm}$; (b) fixed detuning $\delta=-0.13\,\text{nm}$ and input powers $P_1=2.25\,\mu\text{W}$, $P_2=2.35\,\mu\text{W}$, $P_3=2.4\,\mu\text{W}$, $P_4=2.5\,\mu\text{W}$, $P_5=2.6\,\mu\text{W}$, $P_6=2.8\,\mu\text{W}$. }
\label{figure4}
\end{center}
\end{figure*}

We consider an optical resonator in which the optical intensity and the intra-cavity optical path are nonlinearly coupled through a TO effect and a slower thermo-mechanical process of opposite sign. When light is injected into the cavity on the red-side with respect to the initial resonance (with no field), the temperature of the nanobeam changes due to residual optical absorption. This results in a red-shift of the optical resonance via TO effect and thus in an increase of the intra-cavity intensity. The resonant field has the additional effect of slowly blue-shifting the cavity resonance through a thermo-optically induced mechanical deformation. 

Since the optical field evolves on a fast timescale as compared to the thermal effects, it will instantaneously adapt to any change of the resonant condition and thus its dynamics can be adiabatically eliminated.
The model thus reduces to the following system of ODEs:
\begin{eqnarray}
\dot{\phi} &=& - \phi + g_{\phi} I_c (\phi,\theta) \;  \label{eq1} \\
\dot{\theta} &=& -\varepsilon [\theta + g_{\theta} I_c(\phi,\theta)]  \;  \label{eq2} \\
I_c(\phi,\theta) &=& \frac{1}{1 + (\delta_0 + \phi + \theta)^2} \; , \nonumber
\end{eqnarray}
where $\phi$ and $\theta$ describe the instantaneous changes of the resonant wavelength due to thermo-optical and thermo-mechanical effects, respectively; $I_c$ is the intra-cavity field intensity, normalized to its resonant value $I_c^{max}$ and $\delta_0$ is the normalized detuning between laser and cavity resonance. All these quantities are normalized to the cavity half-linewidth (in lenght units) $\gamma$ and are thus dimensionless variables. The time derivatives are calculated with respect to dimensionless time $\gamma_{\rm{to}} t$, where 
$\gamma_{\rm{to}}= G /(C \rho V_\text{c})$ is the "thermo-optical rate": here, G is the thermal conductance between the nanobeam and the substrate, which depends on the thermal conductivity $\kappa$ of Si$_3$N$_4$ and on the geometrical details of the structure, whereas $C$ and $\rho$ are the specific heat capacity and the density of the material, respectively. The dimensionless parameter 
\begin{equation}
g_{\phi}=2\alpha\frac{dn}{dT}\frac{V_\text{c}}{G}\frac{Q}{n_0}I_c^{max}\; \label{eq3}
\end{equation}
measures the strength of the TO effect, where $\alpha$ is the optical-absorption coefficient, $dn/dT$ is the thermo-optical coefficient, $n_0$ is the refractive index, $Q$ is the cavity quality factor and $V_\text{c}$ is the cavity volume. The intra-cavity intensity at resonance $I_c^{max}$ contains the dependency on the input power, and scales as $I_c^{max}= Q \sqrt{T}\,P_\text{in}/A$, where $T$ is the cavity transmittance and $A$ the nanobeam cross section. 

In Eq. (\ref{eq2}), the dimensionless parameter $\varepsilon$ is the ratio between the characteristic rate of the thermo-mechanical effect and $\gamma_{\rm{to}}$, thus $\varepsilon \ll 1$. The thermomechanic parameter $g_{\theta}$ should depend on the thermal expansion coefficient, thermal stresses and geometric details of the PhC nano-structure. Here we phenomenologically express it in terms of the TO strength, as $g_{\theta}=b g_{\phi}$, where $b$ is a positive constant.

The phase space structure of Eqs. (\ref{eq1}-\ref{eq2}) is similar to that of the model in \cite{marino2006,marino2007} describing relaxation oscillations in optical cavities due to competing radiation-pressure and photothermal displacement. Mathematically, it could be derived from Eqs. ($9$) of \cite{marino2006} after adiabatic elimination of the second-order time derivative, i.e. in the singular limit $Q=0$. As such, we expect that many relevant features of that model, in particular the exhibit of an SSP dynamics similar to the VdPFN equations, should be found also in our case. This is what we show in the next section.

As a first check of our approach, we characterize the response of system (\ref{eq1}-\ref{eq2}) as the detuning parameter $\delta_0$ is scanned over the cavity resonance for different injected powers. 
The detuning is scanned at a rate equal to $\varepsilon$, i.e. slower than the TO rate which is $\mathcal{O}(1)$, and comparable to the TM one. The resulting spectra of the intra-cavity intensity, plotted in Fig. \ref{figure3}(d), are in good agreement with the experimental data, while at slower scanning rates the dynamical instability would become manifest in the form of sharp pulsations on the red-side of the resonance. 

\section{Self-sustained pulsations}

We now analyze in detail the dynamical regimes. Fig. \ref{figure4}(a) shows six traces of the transmitted intensity as the detuning between the cavity resonance and the laser frequency is delicately decreased, approaching the resonance from the red side. As we will see in the next section, the steady intensity state becomes unstable in correspondence of a critical value of the detuning via a supercritical Hopf bifurcation, beyond which a finite-frequency limit cycle starts to grow. The pulsation pattern, consisting of fast transitions between high and low intensity states on which the evolution is much slower, is indicative of multiple timescale dynamics.
As the detuning is further decreased, the duty cycle of the oscillation continuously changes and finally a high-intensity steady state is reached. This sequence of dynamical regimes is fully compatible with the transition between the two stable branches of the critical manifold of the vdPFN model \cite{fhnbook}.

A similar behaviour is found by fixing the detuning and progressively increasing the input power, although in this case we also observe an increase in the pulsation period (see Fig. \ref{figure4}(b)). The change in the periodicity is accompanied by a slope decrease in the evolution of the higher-intensity state.

We now compare the model predictions with the above results by numerically solving Eqs.(\ref{eq1}-\ref{eq2}).
The model contains a number of physical parameters that can be measured independently or estimated from the experimental time-series. The switching between the lower and upper transmission states approximately occurs with a characteristic time given by the inverse thermo-optical rate $\tau_\text{to}=1/\gamma_{\rm{to}}$. In our case this time is of the order of $\tau_\text{to}\sim 10\,\text{ms}$, being roughly independent of the values of input power and detuning. We can thus use the measured $\tau_\text{to}$ to estimate the thermal conductivity $\kappa$ of our PhC structure. Assuming that the heat generated by the intra-cavity optical intensity can only diffuse through the thin bridges of the nanobeam (radiative loss of heat is neglected), it is possible to derive an approximate expression for the thermal conductance $G=\frac{\kappa l}{4 e h}$, where $\kappa$ is the thermal conductivity \cite{Haret2009}. Using the values $C=700 \,\rm{J/kg\,K}$ and $\rho=3100 \,\rm{kg/m^3}$ for the specific heat capacity and the density of Si$_3$N$_4$, the cavity volume $V_\text{c}=7.2\times 10^{-18}\rm{m^3}$, and the above $\tau_\text{to}$, we calculate $\kappa \simeq 0.6\,\text{W/m K}$.
Typical values of thermal conductivity for Si$_3$N$_4$ membranes range from $2$ to $4\,\text{W/m K}$ \cite{griffin,sultan,ftouni} although they drop for thicknesses below $200$ nm due to predominant phonon-boundary scattering, in which case also values around $0.5\,\text{W/m\,K}$ have been reported \cite{lee}. 

The parameter $g_{\phi}$ can be experimentally estimated from the transmission spectra in Fig. \ref{figure3}, measuring the thermo-optical shift per unit input power $\partial \phi / \partial P_{\rm{in}}$. From Eq. \ref{eq1} one can readily verify that:
\begin{equation}
\partial \phi / \partial P_{\rm{in}}=g_{\phi}/ P_{\rm{in}}= 2\alpha\frac{dn}{dT}\frac{V_{\rm{m}}}{G}\frac{Q^2}{n_0}\sqrt{T}/A\; ,
\label{shift}
\end{equation}
where we used the relation between resonant intra-cavity intensity and $P_{\rm{in}}$. We obtain a shift of $\sim 8 \times 10^{5}\,\rm{W^{-1}}$ and we will use this value in all the simulations. 
From the shift we can also evaluate the product $\alpha \times \frac{dn}{dT}$. Using the thermal conductance derived from the rise-time measurements, a transmittance $T=0.01$, and the cavity parameters previosly reported, we find $\alpha \times \frac{dn}{dT} \simeq 2 \times 10^{-5} \rm{m^{-1} K^{-1}}$, which quantifies the thermo-optical response in our PhC structure.

\begin{figure}
\begin{center}
\includegraphics[width=1.03\columnwidth]{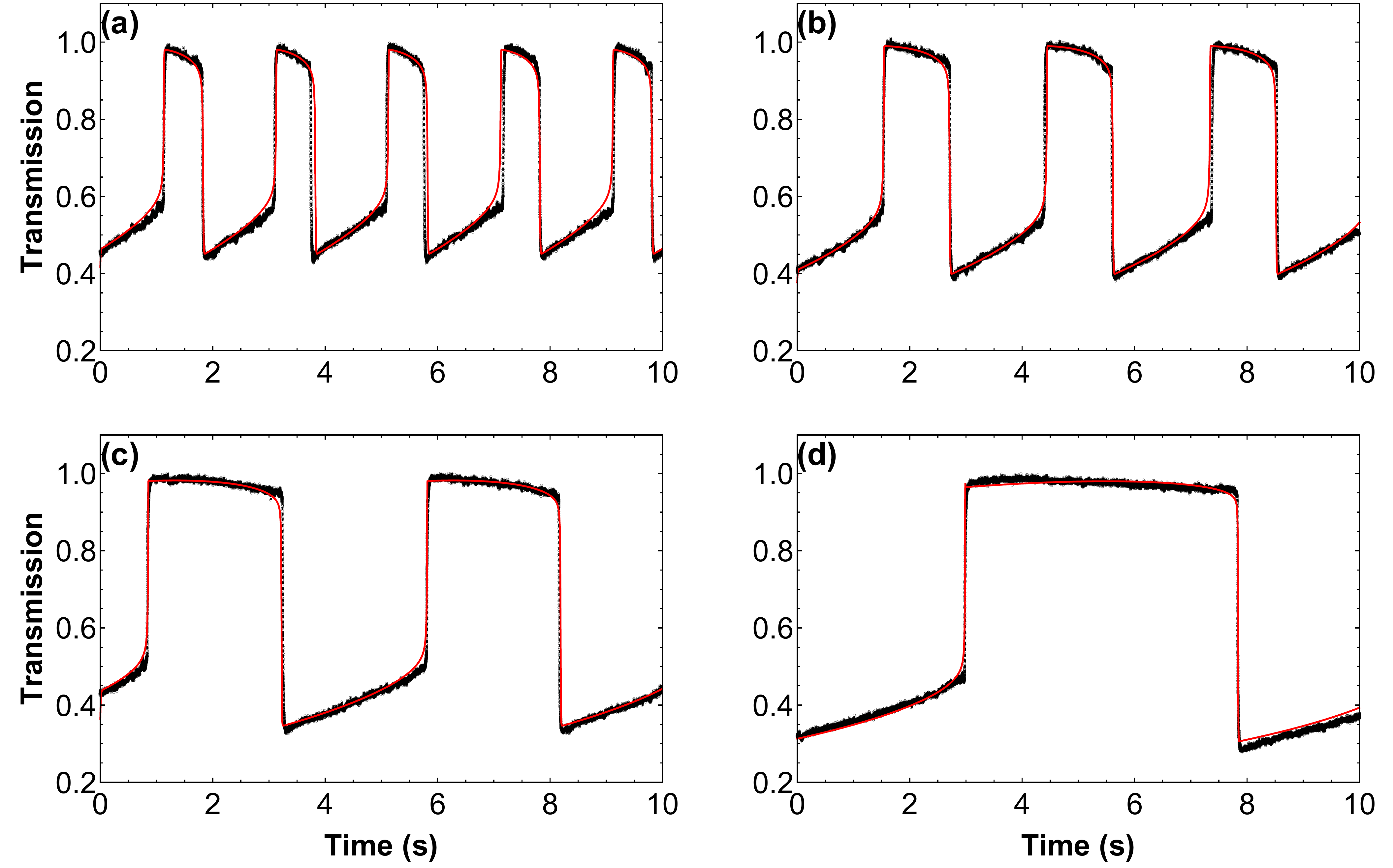}
\caption{Comparison between experimental (black) and numerical time-traces (red) of the optical intensity. Simulations are obtained using the normalized detuning $\delta_0$ given by experimental parameters $\delta/\gamma \sim -1.08$, and an input power $P_\text{in}$ given by (a) $\text{P}_2=2.25\,\mu\text{W}$, (b) $\text{P}_3=2.35\,\mu\text{W}$, (c) $\text{P}_4=2.5\,\mu\text{W}$, and (d) $\text{P}_5=2.8\,\mu\text{W}$ .}
\label{figure5}
\end{center}
\end{figure}

The comparison between numerical and experimental timeseries is illustrated in Fig. \ref{figure5}. In the simulations we fix all thermo-optical parameters and the experimental values of detuning and input powers, and we adjust the phenomenological thermo-mechanical coefficients to match the period, duty cycle and the shape of the SSPs. An excellent agreement is obtained for $b\simeq 0.7$ and $\varepsilon\simeq 3 \times 10^{-3}$.

In the next section we demonstrate that all these features can be explained in terms of the existence of a 1D slow manifold which determines the SSP regime.

\section{Bifurcation analysis}

The steady state solutions of Eqs. (\ref{eq1})-(\ref{eq2}) are implicitly defined by the cubic equation for the stationary intra-cavity intensity, $I_s$:
\begin{equation}
I_s [1 + ( \delta_0 + (g_\phi - g_\theta) I_s)]^2 - 1 = 0 \; .
\label{eq4}
\end{equation}
Depending on the values of $g_\phi$, $g_\theta$, $P_\text{in}$ and $\delta_{0}$, the system can have either one or three fixed points. The change in the number of stationary points occurs when:
\begin{equation}
\left ( \frac{\delta_0^2 - 3}{9} \right )^3 = \left [ \frac{g_\phi - g_\theta}{2} + \frac{1}{3} \delta_0 \left( 1 + \frac{\delta_0^2}{9} \right) \right ]^2 \; .
\label{eq5}
\end{equation}
where two steady states coalesce in a saddle-node bifurcation. Eq. \ref{eq5} thus defines the boundaries in the ($\delta_0$, $P_{\rm{in}}$) parameter space of the region where the system is bistable. These boundaries meet in two cusp points at $\delta_0 = \pm \sqrt{3}$ and $(g_\phi - g_\theta) = \pm 8 / 3 \sqrt{3}$ where a pitch-fork bifurcation takes place (in our case $g_\phi > g_\theta$ and thus the relations with positive signs hold). The second condition thus defines the optical bistable threshold, i.e. the minimum power $P_{\rm{th}}$ at which the system admits two stable states. Using relation (\ref{shift}), the experimental thermo-optical shift $\sim 8 \times 10^{5} \rm{W^{-1}}$, and the fact that $g_\theta \approx 0.7 g_\phi$, we calculate $P_\text{th} \approx 6.5 \mu \rm{W}$.

We now study the stability of the system in the parameter range for which it has a single steady state. It is in this regime that SSPs arise. Linearizing Eqs. (\ref{eq1}-\ref{eq2}) around the fixed point ($\phi_s$, $\theta_s$) we get the following characteristic equation for the eigenvalues $\Lambda$:
\[
\Lambda^2 + a_1 \Lambda + a_2 = 0 \; .
\]
The coefficients are given by:
\begin{eqnarray}
a_1 &=& 1-g_\phi I'_c + \varepsilon (1 + g_\theta I'_c)\; , \nonumber \\
a_2 &=& \varepsilon [1 -(g_\phi-g_\theta)I'_c]  \; ,\nonumber \\
\end{eqnarray}
where $I'_{\rm{c}}=d I_{\rm{c}}(\phi_s,\theta_s)/d (\phi_s+\theta_s)$.
For parameters such that $a_1=0$, the characteristic equation has two purely imaginary roots $\Lambda_{1,2} = \pm i \nu$. Here the steady state loses stability through a supercritical Hopf bifurcation and a quasi-harmonic limit cycle develops, with an amplitude that scales as the square root of the distance from the bifurcation point and frequency given by $\nu=\sqrt{a_2}$. Similarly to the VdPFN equations, the frequency $\nu$ scales as $\sqrt{\varepsilon}$.
However, the large split between time scales associated to the smallness of $\varepsilon$ makes the Hopf limit cycle observable only within a parameter range of order $\varepsilon$ around the bifurcation point. Outside this range the amplitude of the limit cycle abruptly (though continuously) jumps and reaches a saturation value, the so-called relaxation-oscillation regime. Likewise, the frequency of the oscillations experiences a similar sudden change and becomes of the order of $\varepsilon$. Further increasing of $\delta_0$ leads to the “inverse” bifurcation and the system passes from the oscillatory dynamics to a new steady state solution.

\begin{figure*}
\begin{center}
\includegraphics[width=2.0\columnwidth]{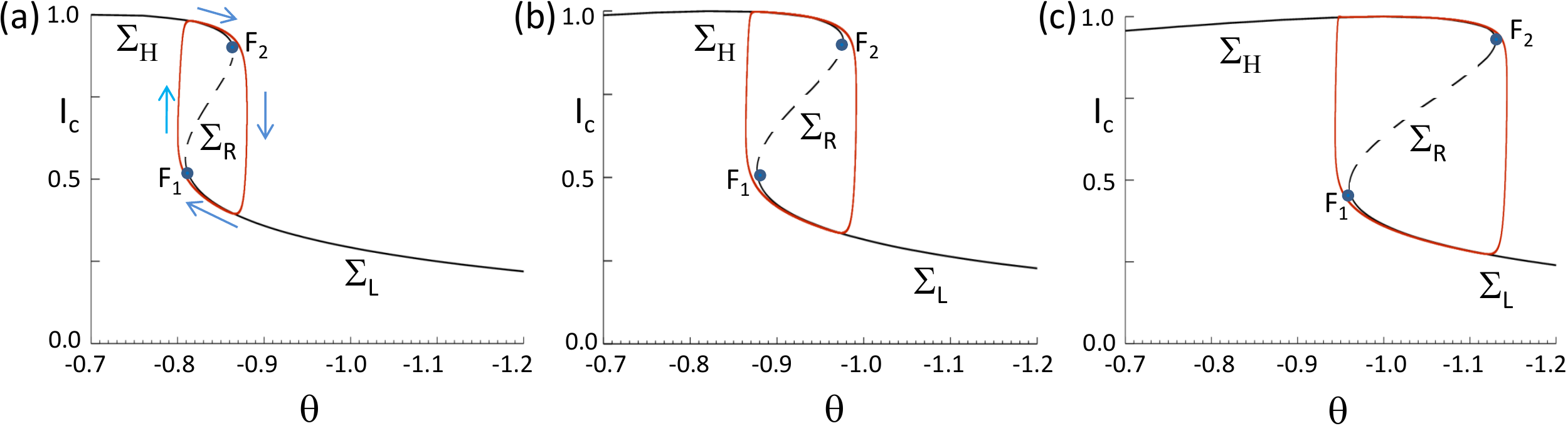}
\caption{Numerical phase-space trajectories (red curves) together with the critical manifold (black curve) Eq. \ref{sm} in the ($I_c$,$\theta$) plane for $\delta_0=1.08$ (a) $P_\text{in} =2.25\,\mu\text{W}$, (b) $P_\text{in}=2.5\,\mu\text{W}$, (c) $P_\text{in}=2.8\,\mu\text{W}$. Solid (dashed) curves indicate the stable (unstable) branches $\Sigma_{H,L}$ ($\Sigma_R$) of the manifold coalescing at the fold points $F_{1,2}$ (see text).}
\label{figure6}
\end{center}
\end{figure*}

\section{Geometric theory of singular perturbation}

The dynamical mechanism underlying the SSP dynamics can be understood by means of the following analysis.
Since $\epsilon \ll 1$, the variable $\theta$ evolves at a much slower rate than $\phi$. Hence the dynamics of Eqs. (\ref{eq1}-\ref{eq2}) splits into periods of fast and slow motion that can be analyzed separately \cite{hirsch1974}. On the fast time scale $t$, the evolution is described by the thermo-optic equation \ref{eq1} (fast subsystem), with $\theta$ acting as a constant parameter. The equilibria of this dynamical subsystem lay on the one-dimensional manifold $\Sigma=\{\phi_s, \theta\}$, implicitly defined by the equation $\phi_s = g_{\phi} I_c(\phi_s,\theta)$ or, equivalently by the cubic:
\begin{equation}
I_c [1 + (\delta_0 + g_{\phi} I_c + \theta)^2] = 0 \; .
\label{sm}
\end{equation}
On the slow time scale $\tau_{\varepsilon}=\varepsilon t$, the motion is governed by the thermo-optomechnical equation (\ref{eq2}) with an algebraic constraint given by $\dot{\phi}=0$ or, equivalently, Eq. (\ref{sm}). Therefore, the (slow) motion on a timescale $\tau_{\varepsilon}$ takes place on the critical manifold $\Sigma$ defined by the fixed points of the fast subsystem. Since the trajectories of Eqs. (\ref{eq1}-\ref{eq2}) will be attracted by stable parts of $\Sigma$, while they will be repelled by the unstable ones \cite{fenichel}, the stability properties of the critical manifold determine the dynamics. Linearizing the fast subsystem on $\Sigma$ we find that these points are stable equilibria if $g_{\phi} I_c(\phi_s,\theta) < 1$ (solid lines in Fig. \ref{figure6}) and unstable otherwise (dashed line). Therefore the critical manifold is composed of two attractive branches of high-intensity $\Sigma_{H}$ and low-intensity $\Sigma_{L}$ states, separated by the repelling branch $\Sigma_R$. Stable and unstable branches coalesce in saddle-node bifurcations at the fold points $F_{1,2}$, which are determined by Eq. (\ref{sm}) together with the condition $g_{\phi} I_c(\phi_s,\theta) < 1$.

We can now understand the blowup of SSPs in our system. In Fig. \ref{figure6} we plot the numerical phase-space trajectories together with the critical manifold $\Sigma$. Depending on the initial conditions, the motion is attracted by either $\Sigma_H$ or $\Sigma_L$. On these branches Eq. (\ref{eq2}) dictates that $\theta$ decreases if $\theta + g_{\theta} I_c(\theta)>0$ and increases otherwise. These conditions determine the flow direction on the slow manifold as indicated by the arrows in Fig. \ref{figure6}. The trajectories are thus forced to (slowly) follow an attracting part of the manifold until the corresponding fold point where it is rapidly pushed out towards the opposite attracting branch. Then, it flows along this branch until the other fold point where it jumps back and repeats the cycle. The two-time scale evolution, slow on the attractive branches $\Sigma_H$ or $\Sigma_L$ and fast in the transitions between them, determine the typical square-wave-like profile of SSPs. The shape of the critical manifold, and the portion of the branches explored by the limit cycle depends on the thermo-optical properties of the PhC cavity, the detuning and the input power through the parameter $g_{\phi}$. In particular, we observe that the width of the slowly evolving parts and the slope of the high-intensity branch $\Sigma_H$ change with the input power, which explains the behaviour observed in Figs. \ref{figure4} and \ref{figure5}.

The bifurcations and the properties of the critical manifold described above are common to many 2D dynamical systems dysplaying relaxation-oscillations and, in particular, to the VdPFN neuron model.

\section{Conclusions and Perspectives}

We have studied the nonlinear optical response of suspended 1D PhC nanocavity devices, fabricated on a Si$_3$N$_4$ chip. Owing to the strong light and heat confinement, thermo-optical nonlinearities become significant at injected powers as low as $\sim2\,\mu\text{W}$. When the laser is detuned to the red-side of the cavity resonance, we observe SSPs of the cavity transmitted signal with sub-Hz periodicity. The observed SSPs are sensitive to small changes in input power and laser wavelength, not just in their period, but also in the duty cycle and oscillations shape. These dynamics are attributed to the interplay between a faster thermo-optical effect and a slower thermo-optomechanical mechanism. On this basis we constructed a simple physical model that reproduces all the observed phenomenology and allows us to evaluate from the time-series relevant quantities to our nanocavities, such as the thermal conductivity and the product $\alpha \times \frac{dn}{dT}$. By means of singular perturbation analysis we have shown that the phase space structure of the system is equivalent to that of the VdPFN model and that all features of SSPs can be explained in terms of the stability properties of a 1D critical manifold on which the slow dynamics takes place. In the vicinity of the Hopf bifurcation point, the system is expected to display excitable features: time-localized perturbations above a certain threshold induce large excursions in the phase space, which are barely sensitive to the details of the perturbation, before returning to the initial state. Excitability is one of the most important functional properties of neurons and photonic systems have long served as a platform for the exploration of this phenomenon \cite{NPreview}. The possibility to couple several nanocavities within a single membrane [see e.g. Fig. \ref{figure1}(a)] thus opens interesting perspectives in the context of neuromorphic photonics, for instance in the implementation of networks of individually addressable, excitable elements. Overall, this work shows that the optical properties of free standing nanophotonic structures can only be understood once their thermal and mechanical characteristics are taken into acount. Therefore, our results give important insights into the design of free standing nanophotonic structures that are actively explored for quantum information processing with individual atoms \cite{Kimble2020}.

\section*{ACKNOWLEDGMENTS}
The authors acknowledge financial support from the European Research Council through Grant QnanoMECA (CoG-64790), Fundació Privada Cellex, CERCA Programme / Generalitat de Catalunya, and the Spanish Ministry of Economy and Competitiveness through the Severo Ochoa Programme for Centres of Excellence in R\&D.\\
P.Z.G. Fonseca and I. Alda contributed equally to this work.

\bibliography{references}

\end{document}